     
\font\twelverm=cmr10 scaled 1200    \font\twelvei=cmmi10 scaled 1200
\font\twelvesy=cmsy10 scaled 1200   \font\twelveex=cmex10 scaled 1200
\font\twelvebf=cmbx10 scaled 1200   \font\twelvesl=cmsl10 scaled 1200
\font\twelvett=cmtt10 scaled 1200   \font\twelveit=cmti10 scaled 1200
     
\skewchar\twelvei='177   \skewchar\twelvesy='60
     
     
\def\twelvepoint{\normalbaselineskip=12.4pt
  \abovedisplayskip 12.4pt plus 3pt minus 9pt
  \belowdisplayskip 12.4pt plus 3pt minus 9pt
  \abovedisplayshortskip 0pt plus 3pt
  \belowdisplayshortskip 7.2pt plus 3pt minus 4pt
  \smallskipamount=3.6pt plus1.2pt minus1.2pt
  \medskipamount=7.2pt plus2.4pt minus2.4pt
  \bigskipamount=14.4pt plus4.8pt minus4.8pt
  \def\rm{\fam0\twelverm}          \def\it{\fam\itfam\twelveit}%
  \def\sl{\fam\slfam\twelvesl}     \def\bf{\fam\bffam\twelvebf}%
  \def\mit{\fam 1}                 \def\cal{\fam 2}%
  \def\tt{\twelvett}
  \textfont0=\twelverm   \scriptfont0=\tenrm   \scriptscriptfont0=\sevenrm
  \textfont1=\twelvei    \scriptfont1=\teni    \scriptscriptfont1=\seveni
  \textfont2=\twelvesy   \scriptfont2=\tensy   \scriptscriptfont2=\sevensy
  \textfont3=\twelveex   \scriptfont3=\twelveex  \scriptscriptfont3=\twelveex
  \textfont\itfam=\twelveit
  \textfont\slfam=\twelvesl
  \textfont\bffam=\twelvebf \scriptfont\bffam=\tenbf
  \scriptscriptfont\bffam=\sevenbf
  \normalbaselines\rm}
     
     
\def\tenpoint{\normalbaselineskip=12pt
  \abovedisplayskip 12pt plus 3pt minus 9pt
  \belowdisplayskip 12pt plus 3pt minus 9pt
  \abovedisplayshortskip 0pt plus 3pt
  \belowdisplayshortskip 7pt plus 3pt minus 4pt
  \smallskipamount=3pt plus1pt minus1pt
  \medskipamount=6pt plus2pt minus2pt
  \bigskipamount=12pt plus4pt minus4pt
  \def\rm{\fam0\tenrm}          \def\it{\fam\itfam\tenit}%
  \def\sl{\fam\slfam\tensl}     \def\bf{\fam\bffam\tenbf}%
  \def\smc{\tensmc}             \def\mit{\fam 1}%
  \def\cal{\fam 2}%
  \textfont0=\tenrm   \scriptfont0=\sevenrm   \scriptscriptfont0=\fiverm
  \textfont1=\teni    \scriptfont1=\seveni    \scriptscriptfont1=\fivei
  \textfont2=\tensy   \scriptfont2=\sevensy   \scriptscriptfont2=\fivesy
  \textfont3=\tenex   \scriptfont3=\tenex     \scriptscriptfont3=\tenex
  \textfont\itfam=\tenit
  \textfont\slfam=\tensl
  \textfont\bffam=\tenbf \scriptfont\bffam=\sevenbf
  \scriptscriptfont\bffam=\fivebf
  \normalbaselines\rm}
     
     
\def\beginlinemode{\endmode
  \begingroup\parskip=0pt \obeylines\def\\{\par}\def\endmode{\par\endgroup}}
\def\beginparmode{\endmode
  \begingroup \def\endmode{\par\endgroup}}
\let\endmode=\par
{\obeylines\gdef\
{}}
\def\singlespace{\baselineskip=\normalbaselineskip}

\def\oneandahalfspace{\baselineskip=\normalbaselineskip
  \multiply\baselineskip by 3 \divide\baselineskip by 2}
\def\doublespace{\baselineskip=\normalbaselineskip \multiply\baselineskip by 2}

\newcount\firstpageno
\firstpageno=2
\footline={\ifnum\pageno<\firstpageno{\hfil}\else{\hfil\twelverm\folio\hfil}\fi}
\let\rawfootnote=\footnote              
\def\footnote#1#2{{\tenrm\singlespace\parindent=0pt\rawfootnote{#1}{#2}}}
\def\raggedcenter{\leftskip=4em plus 12em \rightskip=\leftskip
  \parindent=0pt \parfillskip=0pt \spaceskip=.3333em \xspaceskip=.5em
  \pretolerance=9999 \tolerance=9999
  \hyphenpenalty=9999 \exhyphenpenalty=9999 }
\def\dateline{\rightline{\ifcase\month\or
  January\or February\or March\or April\or May\or June\or
  July\or August\or September\or October\or November\or December\fi
  \space\number\year}}
\def\received{\vskip 3pt plus 0.2fill
 \centerline{\sl (Received\space\ifcase\month\or
  January\or February\or March\or April\or May\or June\or
  July\or August\or September\or October\or November\or December\fi
  \qquad, \number\year)}}
     
     
\hsize=6.5truein
\hoffset=0truein
\vsize=8.9truein
\voffset=0truein
\parskip=\smallskipamount
\twelvepoint            
\doublespace            
\overfullrule=0pt       
     
     
\def\preprintno#1{
 \rightline{\rm #1}}    
     
\def\title                      
  {\null\vskip 3pt plus 0.2fill
   \beginlinemode \doublespace \raggedcenter \bf}
     
\def\author                     
  {\vskip 3pt plus 0.2fill \beginlinemode
   \singlespace \raggedcenter}
     
\def\affil                      
  {\vskip 3pt plus 0.1fill \beginlinemode
   \oneandahalfspace \raggedcenter \sl}
     
\def\abstract                   
  {\vskip 3pt plus 0.3fill \beginparmode
   \doublespace \narrower ABSTRACT: }
     
\def\endtitlepage               
  {\endpage                     
   \body}
     
\def\body                       
  {\beginparmode}               
     
\def\head#1{                    
  \filbreak\vskip 0.5truein     
  {\immediate\write16{#1}
   \raggedcenter \uppercase{#1}\par}
   \nobreak\vskip 0.25truein\nobreak}

\def\refto#1{$|{#1}$}           
     
\def\references                 
  {\head{References}            
   \beginparmode
   \frenchspacing \parindent=0pt \leftskip=1truecm
   \parskip=8pt plus 3pt \everypar{\hangindent=\parindent}}
     
\gdef\refis#1{\indent\hbox to 0pt{\hss#1.~}}    
     
\gdef\journal#1, #2, #3, 1#4#5#6{               
    {\sl #1~}{\bf #2}, #3, (1#4#5#6)}           
     
\def\refstylenp{                
  \gdef\refto##1{ [##1]}                                
  \gdef\refis##1{\indent\hbox to 0pt{\hss##1)~}}        
  \gdef\journal##1, ##2, ##3, ##4 {                     
     {\sl ##1~}{\bf ##2~}(##3) ##4 }}
     
\def\refstyleprnp{              
  \gdef\refto##1{ [##1]}                                
  \gdef\refis##1{\indent\hbox to 0pt{\hss##1)~}}        
  \gdef\journal##1, ##2, ##3, 1##4##5##6{               
    {\sl ##1~}{\bf ##2~}(1##4##5##6) ##3}}

\def\figurecaptions             
  {\endpage
   \beginparmode
   \head{Figure Captions}
}

\def\endpage                    
  {\vfill\eject}
     
\def\endpaper                   
  {\endmode\vfill\supereject}

     
\def\ref#1{Ref. #1}                     

\def\frac#1#2{{\textstyle{#1 \over #2}}}

\def\sla{\raise.15ex\hbox{$/$}\kern-.57em}
\def\leaderfill{\leaders\hbox to 1em{\hss.\hss}\hfill}
\def\twiddle{\lower.9ex\rlap{$\kern-.1em\scriptstyle\sim$}}
\def\bigtwiddle{\lower1.ex\rlap{$\sim$}}
\def\gtwid{\mathrel{\raise.3ex\hbox{$>$\kern-.75em\lower1ex\hbox{$\sim$}}}}
\def\ltwid{\mathrel{\raise.3ex\hbox{$<$\kern-.75em\lower1ex\hbox{$\sim$}}}}
\def\square{\kern1pt\vbox{\hrule height 1.2pt\hbox{\vrule width 1.2pt\hskip 3pt
   \vbox{\vskip 6pt}\hskip 3pt\vrule width 0.6pt}\hrule height 0.6pt}\kern1pt}

\def\Fint{\rlap{$\Biggl\rfloor$}\Biggl\lceil}

\def\m@th{\mathsurround=0pt }
\def\leftrightarrowfill{$\m@th \mathord\leftarrow \mkern-6mu
 \cleaders\hbox{$\mkern-2mu \mathord- \mkern-2mu$}\hfill
 \mkern-6mu \mathord\rightarrow$}
\def\overleftrightarrow#1{\vbox{\ialign{##\crcr
     \leftrightarrowfill\crcr\noalign{\kern-1pt\nointerlineskip}
     $\hfil\displaystyle{#1}\hfil$\crcr}}}

\singlespace
\preprintno{gr-qc/9803096}
\preprintno{UFIFT-HEP-98-3}
\vskip 2cm
\centerline{\bf PARTICLES AS BOUND STATES IN THEIR OWN POTENTIALS$^{\dagger}$}
\vskip 1cm
\centerline{\bf R. P. Woodard}
\vskip .5cm
\centerline{\it Department of Physics}
\centerline{\it University of Florida}
\centerline{\it Gainesville, FL 32611}
\vskip 1cm
\centerline{\tenrm ABSTRACT}
\itemitem{}{\tenrm I consider the problem of computing the mass of a charged, 
gravitating particle in quantum field theory. It is shown how solving for the 
first quantized propagator of a particle in the presence of its own potentials 
reproduces the gauge and general coordinate invariant sum over an infinite 
class of diagrams. The distinguishing feature of this class of diagrams is that
all closed loops contain part of the continuous matter line running from early 
to late times. The next order term would have one closed loop external to the
continuous matter line, and so on. I argue that the gravitational potentials in
the 0-th order term may permit the formation of bound states, which would then 
dominate the propagator. It is conceivable that this provides an tractable 
technique for computing the masses of fundamental particles from first 
principles. It is also conceivable that the expansion in external loops permits
gravity to regulate certain ultraviolet divergences.}

\footnote{}{$^{\dagger}$ To appear in the proceedings of {\tenit Physics Of 
Mass}, Miami Beach, Dec. 12-15, 1997}

\line{\hfil}
\line{\hfil}
\line{\hfil}
\line{\bf 1. INTRODUCTION: A PARABLE OF POLITICAL CORRECTNESS \hfil}
\line{\hfil}
\indent I wish to speak out against a form of bigotry. The prejudice in 
question might be termed, {\it integro-centrism}, and it consists of the belief
that asymptotic series may contain only non-negative, integer powers of the 
expansion coefficient. Not only is this exclusionary against non-integer powers
and logarithms, it even discrimates against sign-challenged integers! I shall 
also argue that integro-centrism may be imposing a kind of cultural genocide on 
quantum gravity and on the problem of mass.\hfil\break
\indent Imagine that you are the asymptotic expansion ${\widetilde f}(g)$ of
some quantum field theoretic quantity $f(g)$. Without succumbing to negative 
stereotypes we can assume you have the following form:
$${\widetilde f}(g) = \sum_{n=0}^{\infty} f_n \phi_n(g) \; . \eqno(1.1)$$
We can also assume that the $\phi_n(g)$ are elementary functions which have 
been arranged in a (value-neutral) order such that:
$$\lim_{g \rightarrow 0} {\phi_{n+1}(g) \over \phi_n(g)} = 0 \; . \eqno(1.2)$$
Finally, the fact that you are asymptotic means that the difference between 
$f(g)$ and the sum of your first $N$ terms must vanish faster than your $N$-th 
term as $g$ goes to zero:
$$\lim_{g \rightarrow 0} \Bigl(f(g) - \sum_{n=0}^N f_n \phi_n(g)\Bigr) 
{1 \over \phi_N(g)} = 0 \; . \eqno(1.3)$$
\indent In the Ward and June Cleaver world of conventional perturbation theory 
the coefficient functions would be integer powers --- $\phi_n(g) = g^n$ --- and
their coefficients could be obtained by taking derivatives of the original 
function $f(g)$:
$$f_n = {1 \over n!} {\partial^n f(g) \over \partial g^n} \Bigl\vert_{g = 0} 
\; . \eqno(1.4)$$
Suppose, however, that you are leading an alternate lifestyle which includes 
logarithms or fractional powers. For example, you might have the form:
$${\widetilde f}(g) = 1 + 3 g \ln(g) + O(g^2) \;. \eqno(1.5)$$
Although your actual first order correction is small for small $g$, an 
integro-centric bigot would claim it is logarithmically divergent: 
$$f_1 = \lim_{g \rightarrow 0} \Bigl(3 \ln(g) + 3 + O(g)\Bigr)\; . \eqno(1.6)$$
And he would compute the higher terms to consist of an oscillating tower of
increasingly virulent divergences:
$$f_n = \lim_{g \rightarrow 0} \Bigl(- \frac3{n} (-g)^{1-n} + \dots\Bigr) 
\qquad , \qquad n \geq 2 \; . \eqno(1.7)$$
His frustration with your non-conformism might provoke him to abandon the 
quantum field theory behind $f(g)$ in favor of some yet-to-be-specified model
in a peculiar dimension. He might even take to making optimistic pronouncements
about our ability to exactly solve this model, and hence its correspondence 
limits of Yang-Mills and General Relativity, in 5-10 years (3-8 years from now,
and counting).
\hfil\break
\indent Aside from poking fun at the political and scientific prejudices of my
colleagues this paper does have some serious points to make. The first of these
is that there is no reason why the perturbative non-renormalizability$^{1-5}$ 
of General Relativity necessarily implies the need for an alternate theory of 
quantum gravity. It has long been realized that the problem could derive 
instead from the appearance of logarithms or fractional powers of Newton's 
constant in the correct asymptotic expansion of quantum gravity.$^{6-9}$ To
underscore that this would not be without precedent I devote Section 2 to a
discussion of the analogous phenomenon in two simple systems from statistical 
mechanics.
\hfil\break
\indent The second point I wish to make is that the breakdown of conventional 
perturbation theory in quantum gravity is likely to be associated with 
ultraviolet divergences. The idea is that gravity screens effects which tend to 
make the stress tensor divergent. If so, it must be that the divergence returns
when Newton's constant goes to zero, which means the correct asymptotic series 
must contain logarithms or negative powers. There is no doubt that this does 
occur on the classical level. Arnowit, Deser and Misner found an explicit 
example in the finite self-energy of point charged particles.$^{10}$ Section 3 
is devoted to a brief review of their result.
\hfil\break
\indent So far I have been discussing old stuff. Although many people have 
suspected that quantum gravity regulates ultraviolet divergences$^{6-9}$ no one
has been able to make anything of the idea for want of a non-perturbative 
calculational technique. Divergences {\it do} evoke an infinite response from
gravitation, but only at the next order in perturbation theory. What is needed
is a way of reorganizing perturbation theory so that the gravitational response
has a chance of keeping up with divergences. The main point of this paper is 
that I have found such a reorganization, at least for the special case of 
certain types of matter self-energies.\hfil\break
\indent I begin the derivation in Section 4 by writing down an exact functional
integral representation for the mass of a charged, gravitating scalar in 
quantum field theory. In Section 5 I show that this expression reduces, in the 
classical limit, to the point particle system studied by ADM,$^{10}$ with an 
extra term representing the negative pressure needed to hold the point charge 
together. In Section 6 I return to the original, exact expression, and show how
it can be rearranged to give an expansion in the number of closed loops which 
do not include a least some part of the incoming and outgoing matter line. 
Further, the 0-th order term in this new expansion has the simple 
interpretation of computing the binding energy of a quantum mechanical particle
which moves in the gravitational and electromagnetic potentials induced by its 
own probability current. This is the origin of the title.\hfil\break
\indent Gravitational attraction must overcome electrostatic repulsion in 
order for a particle to bind to its own potentials. In Section 7 I obtain the 
unsurprising result that this can only happen for a scalar which has a Planck 
scale mass. In the final section I argue that substantially lighter masses may 
be obtainable for particles with spin.\hfil\break
\line{\hfil}
\line{\hfil}
\line{\bf 2. TWO EXAMPLES FROM STATISTICAL MECHANICS \hfil}
\line{\hfil}
\indent Exotic terms occur in many familiar asymptotic expansions. Consider the 
logarithm of the grand canonical partition function for non-interacting, 
non-relativistic bosons in a three dimensional volume $V$:
$$\ln\Bigl(\Xi\Bigr) = V \; n_Q \; \sum_{k=1}^{\infty} k^{-\frac52} \; \exp(k
\beta \mu) \; . \eqno(2.1)$$
Here $n_Q$ is the quantum concentration, $\mu$ is the chemical potential, and 
$\beta = (k_B T)^{-1}$. Near condensation one has $0 < -\beta \mu \ll 1$ so it
should make sense to expand $\ln(\Xi)$ for small $\beta \mu$. Straightforward
perturbation theory corresponds to the following expansion:
$$\eqalignno{\ln\Bigl(\Xi\Bigr) &=  V \; n_Q \; \sum_{k=1}^{\infty} 
k^{-\frac52} \; \sum_{\ell=0}^{\infty} \; (k \beta \mu)^{\ell} &(2.2a) \cr
& \rightarrow  V \; n_Q \; \sum_{\ell=0}^{\infty} \; (\beta \mu)^{\ell} \;
\sum_{k=1}^{\infty} \; k^{\ell-\frac52} \; . &(2.2b) \cr}$$
Although the $\ell=0$ and $\ell=1$ terms are finite, the sum over $k$ diverges
for $\ell \geq 2$. \hfil\break
\indent The divergences we have encountered do not mean that higher corrections
are large, just that they are not as small as $(\beta \mu)^2$. One sees this by
expanding the second derivative around its integral approximation:
$$\eqalignno{{\partial^2 \ln\Bigl(\Xi\Bigr) \over \partial (\beta \mu)^2} \; &= 
\; V \; n_Q \; \sum_{k=1}^{\infty} k^{-\frac12} \; \exp(k \beta \mu) &(2.3a)\cr
&=  V \; n_Q \; \Biggl\{ \int_0^{\infty} dy \; y^{-\frac12} \; \exp(y \beta 
\mu) \cr
& \quad {\hskip 2cm} + \sum_{k=1}^{\infty} \; \Bigl[ k^{-\frac12} \; 
\exp(k \beta \mu) - \int_{k-1}^k dy \; y^{-\frac12} \; \exp(y \beta \mu) 
\Bigr]\Biggr\} \qquad &(2.3b) \cr
&= V \; n_Q \; \Biggl\{ \Bigl({-\pi \over \beta \mu}\Bigr)^{\frac12} + 
\sum_{k=1}^{\infty} \Bigl[k^{-\frac12} - 2 \; k^{\frac12} + 2 \; 
(k-1)^{\frac12}~\Bigr] + O\Bigl(\beta \mu\Bigr) \Biggr\} \; . \qquad 
&(2.3c) \cr}$$
Integration reveals the true asymptotic expansion:
$$\ln\Bigl(\Xi\Bigr) = V \; n_Q \; \Biggl\{\zeta\Bigl(\frac52\Bigr) + 
\zeta\Bigl(\frac32\Bigr) \; \beta \mu + \frac43 \sqrt{\pi} (-\beta \mu)^{
\frac32} + O\Bigl(\beta^2 \mu^2\Bigr)\Biggr\} \; . \eqno(2.4)$$
The oscillating series of ever-increasing divergences in the perturbative
expansion (2.2b) has resolved itself into a perfectly finite, fractional power.
\hfil\break
\indent Logarithms can also invalidate perturbation theory. Consider the 
canonical partition function for a non-interacting particle of mass $m$ in a 
three dimensional volume $V$:
$$\eqalignno{Z &= {V \over 2 \pi^2 \hbar^3} \; \int_0^{\infty} dp \; p^2 \; 
\exp\Bigl[- \beta \sqrt{p^2 c^2 + m^2 c^4} + \beta m c^2\Bigr] &(2.5a) \cr
&= {V \over 2 \pi^2 \hbar^3 c^3} \; \int_0^{\infty} dK \; (K + m c^2) \;
\sqrt{K^2 + 2 \; K m c^2} \; \exp(-\beta K) \; . &(2.5b) \cr}$$
When the rest mass energy is small compared to the thermal energy it ought to
make sense to expand in the small parameter $x \equiv \beta m c^2$. But
straightforward perturbation theory fails again:
$$\eqalignno{Z &= {V \over 2 \pi^2} \; \Bigl({k_B T \over \hbar c}\Bigr)^3 \; 
\int_0^{\infty} dt \; t^2 e^{-t} \; \Bigl(1 + \frac{x}{t}\Bigr) \; \sqrt{1 + 2 
\frac{x}{t}} &(2.6a) \cr
&= {V \over 2 \pi^2} \; \Bigl({k_B T \over \hbar c}\Bigr)^3 \; \int_0^{\infty} 
dt \; t^2 e^{-t} \; \Biggl\{1 + 2 \frac{x}{t} + \frac12 \Bigl(\frac{x
}{t}\Bigr)^2 \cr
& {\hskip 5cm} - \sum_{n=3}^{\infty} {(n-3) \; (2n-5)!! \over n!} \; 
\Bigl(-\frac{x}{t}\Bigr)^n\Biggr\} \; . \; \; &(2.6b) \cr}$$
It seems as though the term of order $x^3$ vanishes, and that the higher terms
have increasingly divergent coefficients with oscillating signs. In fact the
$x^3$ term is non-zero, and the apparent divergences merely signal 
contamination with logarithms:
$$Z  =  {V \over 2 \pi^2} \; \Bigl({k_B T \over \hbar c}\Bigr)^3 \; \Biggl\{2 + 
2 \; x + \frac12 \; x^2 - \frac16 \; x^3 - \frac1{48} \; x^4 \ln(x) + 
O\Bigl(x^4\Bigr)\Biggr\} \; . \eqno(2.7)$$
\line{\hfil}
\line{\hfil}
\line{\bf 3. THE ADM MECHANISM \hfil}
\line{\hfil}
\indent Arnowitt, Deser and Misner showed that perturbation theory also breaks
down in computing the self-energy of a classical, charged, gravitating point 
particle.$^{10}$ It is simplest to model the particle as a stationary spherical 
shell of radius $\epsilon$, charge $e$ and bare mass $m_0$. In Newtonian 
gravity its energy would be:
$$E = m_0 + {e^2 \over 8 \pi \epsilon} - {G m_0^2 \over 2 \epsilon} \; .
\eqno(3.1)$$
It turns out that all the effects of general relativity are accounted for by 
replacing $E$ and $m_0$ with the full mass:\footnote{*}{It should be noted that
Arnowitt, Deser and Misner rigorously solved the constraint equations of 
general relativity and electrodynamics, and then used the asymptotic metric to 
compute the ADM mass.$^{10}$ They also developed the simple model I am 
presenting.}
$$\eqalignno{m_{\epsilon} &= m_0 + {e^2 \over 8 \pi \epsilon} - {G m^2_{
\epsilon} \over 2 \epsilon} &(3.2a) \cr
&= {\epsilon \over G} \Biggl[-1 + \sqrt{1 + {2 G \over \epsilon} \Bigl(m_0 +
{e^2 \over 8 \pi \epsilon}\Bigr)}~\Biggr] \; . &(3.2b) \cr}$$
\indent The perturbative result is obtained by expanding the square root:
$$m_{\rm pert} = m_0 + {e^2 \over 8 \pi \epsilon} + \sum_{n=2}^{\infty} 
{(2n-3)!! \over n!} \; \Biggl(-{G \over \epsilon}\Biggr)^{n-1} \Biggl(m_0 + 
{e^2 \over 8 \pi \epsilon}\Biggr)^n \; , \eqno(3.3)$$
and shows the oscillating series of increasingly singular terms characteristic
of the previous examples. The alternating sign derives from the fact that 
gravity is attractive. The positive divergence of order $e^2/\epsilon$ evokes a
negative divergence or order $G e^4/\epsilon^3$, which results in a positive 
divergence of order $G^2 e^6/\epsilon^5$, and so on. The reason these terms are
increasingly singular is that the gravitational response to an effect at one 
order is delayed to a higher order in perturbation theory.\hfil\break
\indent The correct result is obtained by taking $\epsilon$ to zero before 
expanding in the coupling constants $e^2$ and $G$:
$$\lim_{\epsilon \rightarrow 0} m_{\epsilon} = \Biggl({e^2 \over 4 \pi G}
\Biggr)^{\frac12} \; . \eqno(3.4)$$
Like the examples of Section 2 it is finite but not analytic in the coupling
constants $e^2$ and $G$. Unlike the previous examples, it diverges for small 
$G$. This is because gravity has regulated the linear self-energy divergence 
which results for a non-gravitating charged particle.\hfil\break
\indent One can understand the process from the fact that gravity has a 
built-in tendency to oppose divergences. A charge shell does not want to 
contract in pure electromagnetism; the act of compressing it calls forth a huge
energy density concentrated in the nearby electric field. Gravity, on the other
hand, tends to make things collapse, especially large concentrations of energy
density. The dynamical signature of this tendency is the large negative energy
density concentrated in the Newtonian gravitational potential. In the limit the
two effects balance and a finite total mass results.\hfil\break 
\indent Said this way, there seems no reason why gravitational interactions 
should not act to cancel divergences in quantum field theory. It is especially
significant, in this context, that the divergences of some quantum field 
theories --- such as QED --- are weaker than the linear ones which ADM have 
shown that classical gravity regulates. The frustrating thing is that one 
cannot hope to see the cancellation perturbatively. In perturbation theory the
gravitational response to an effect at any order must be delayed to a higher 
order. This is why the perturbative result (3.3) consists of an oscillating 
series of ever higher divergences. What is needed is an approximation technique
in which the gravitational response is able to keep pace with what is going on
in other sectors.\hfil\break
\indent A final point of interest is that any finite bare mass drops out of the
exact result (3.4) in the limit $\epsilon \rightarrow 0$. This makes for an 
interesting contrast with the usual program of renormalization. Without gravity
one would pick the desired physical mass, $m_p$, and then adjust the bare mass
to be whatever divergent quantity was necessary to give it:
$$m_0 = m_p - {e^2 \over 8 \pi \epsilon} \; . \eqno(3.5)$$
Of course the same procedure would work with gravity as well:
$$m_0 = m_p - {e^2 \over 8 \pi \epsilon} + {G m_p^2 \over 2 \epsilon} \; .
\eqno(3.6)$$
The difference with gravity is that we have an alternative: keep $m_0$ finite
and let the dynamical cancellation of divergences produce a unique result for 
the physical mass. The ADM mechanism is in fact the classical realization of 
the old dream of computing a particle's mass from its self-interactions.
\hfil\break
\line{\hfil}
\line{\hfil}
\line{\bf 4. MASS OF A CHARGED GRAVITATING SCALAR IN QFT \hfil}
\line{\hfil}
\indent The purpose of this section is to obtain a conveneient functional
integral representation for the standard quantum field theoretic definition of
a particle's mass as the pole of its propagator. For simplicity I will consider
a charged, gravitating scalar, the Lagrangian for which is:
$$\eqalignno{{\cal L}= {1 \over 16 \pi G} \Bigl(R &\sqrt{-g} - {\rm S.T.}\Bigr) 
- \frac14 F_{\mu\nu} F_{\rho \sigma} g^{\mu \rho} g^{\nu \sigma} \sqrt{-g} \cr
& - \Bigl(\partial_{\mu} - i q A_{\mu}\Bigr) \phi^* \Bigl(\partial_{\nu} + i q
A_{\nu}\Bigr) \phi \thinspace g^{\mu \nu} \sqrt{-g} - m_0^2 \thinspace \phi^* 
\phi \sqrt{-g} \; . &(4.1a) \cr}$$
The symbol ``S.T.'' denotes the gravitational surface term needed to purge the 
Lagrangian of second derivatives:
$${\rm S.T.} \equiv \partial_{\mu}\Bigl[(g_{\nu \rho , \sigma} - g_{\rho \sigma
, \nu}) g^{\mu \nu} g^{\rho \sigma} \sqrt{-g}\Bigr] \; . \eqno(4.1b)$$
\indent If we temporarily regulate infrared divergences and agree to understand
operator relations in the weak sense then it is possible to write the operators
which annihilate outgoing particles and create incoming ones as simple 
limits:\footnote{*}{The notation employed in these formulae is standard: ${
\scriptstyle Z}$ is the field strength renormalization, the Wronskian is
${\overleftrightarrow {\scriptstyle W}} {\scriptstyle \equiv} {\overrightarrow 
{\scriptstyle \partial}}_{\scriptscriptstyle 0} {\scriptstyle -} 
{\overleftarrow {\scriptstyle \partial}}_{\scriptscriptstyle 0}$, and a tilde 
over the scalar field denotes its spatial Fourier transform.}
$$\eqalignno{a_k^{\rm out} &= \lim_{\scriptscriptstyle t_+ \rightarrow \infty} 
{i e^{i \omega t_+} \over \sqrt{2 \omega Z}} \overleftrightarrow{W}_+ 
{\widetilde \phi}(t_+,{\vec k}) \; , & (4.2a) \cr
\Bigl(a_k^{\rm in}\Bigr)^+ &= \lim_{\scriptscriptstyle t_- \rightarrow -\infty}
{i e^{-i\omega t_-} \over \sqrt{2 \omega Z}} \overleftrightarrow{W}_- 
{\widetilde \phi}^*(t_-,{\vec k}) \; , & (4.2b) \cr}$$
where the energy is:
$$\omega \equiv \sqrt{k^2 + m^2} \; . \eqno(4.3)$$
Consider single particle states whose wave functions in the infinite past and 
future are $\psi_{\mp}$, respectively. The inner product between two such 
states can be given the following expression:
$$\eqalignno{\Bigl\langle \psi_+^{\rm out} \Bigl\vert \psi_-^{\rm in} 
\Bigr\rangle = \int &{d^3 k \over (2 \pi)^3} {\psi_+^*({\vec k}) \psi_-({\vec 
k}) \over 2 \omega Z} \lim_{\scriptscriptstyle t_{\pm} \rightarrow \pm \infty} 
e^{i \omega (t_+ - t_-)} \overleftrightarrow{W}_+ \overleftrightarrow{W}_- \cr
& \times \int d^3x \; e^{-i {\vec k} \cdot {\vec x}} \Bigl\langle \Omega^{\rm 
out} \Bigl\vert \phi(t_+,{\vec x}) \phi^*(t_-,{\vec 0}) \Bigr\vert \Omega^{\rm 
in} \Bigr\rangle \; . &(4.4) \cr}$$
One way of computing the mass is to tune the parameter $m$ in the energy (4.3)
to the precise value for which expression (4.4) assumes the form:
$$\Bigl\langle \psi_+^{\rm out} \Bigl\vert \psi_-^{\rm in} \Bigr\rangle = 
\int {d^3 k \over (2 \pi)^3} {1 \over 2 \omega} \psi_+^*({\vec k}) 
\psi_-({\vec k}) \; . \eqno(4.5)$$
This agrees with the usual definition of the mass as the pole of the 
propagator.\hfil\break
\indent A somewhat more direct way of computing the mass is to focus on the 
second line of (4.4) which we can write as a phase:
$$e^{- i \xi(t_+,t_-,k)} \equiv \int d^3x \; e^{-i {\vec k} \cdot {\vec x}} 
\Bigl\langle \Omega^{\rm out} \Bigl\vert \phi(t_+,{\vec x})\phi^*(t_-,{\vec 0})
\Bigr\vert \Omega^{\rm in} \Bigr\rangle \; . \eqno(4.6)$$
Dividing by the time interval and then taking it to infinity we obtain the 
energy:
$$\lim_{\scriptscriptstyle t_{\pm} \rightarrow \pm\infty} \Bigl({\xi(t_+,t_-,k)
\over t_+ - t_-}\Bigr) = \sqrt{k^2 + m^2} \; . \eqno(4.7)$$
Note that by using this method we avoid the problem of infrared divergences.
These affect only the field strength renormalization, not the mass.\hfil\break
\indent It is straightforward to write the phase as a functional integral:
$$e^{-i \xi} = \int d^3x \; e^{-i {\vec k} \cdot {\vec x}} \Fint [dg] [dA] 
[d\phi] \thinspace \phi(t_+,{\vec x}) \thinspace \phi^*(t_-,{\vec 0})\thinspace
e^{i S_{\rm GR}[g] + i S_{\rm EM}[g,A] + i S_{\phi}[g,A,\phi]}\; . \eqno(4.8)$$
The next step is to integrate out the scalar. In the presence of an arbitrary
metric and electromagnetic background its kinetic operator is:
$${\cal D}[g,A] \equiv {1 \over \sqrt{-g}} \Bigl(\partial_{\mu} + i q A_{\mu}
\Bigr) \sqrt{-g} g^{\mu \nu} \Bigl(\partial_{\nu} + i q A_{\nu} \Bigr) \; .
\eqno(4.9)$$
We can use this operator to express the scalar-induced effective action:
$$\Gamma_{\phi}[g,A] \equiv -i \ln\Bigl( {\rm det}[-{\cal D} + m_0^2 - i 
\epsilon]\Bigr) \; , \eqno(4.10a)$$
and the scalar propagator in the presence of an general background:
$$D[g,A]\Bigl(t_+,{\vec x};t_-,{\vec 0}\Bigr) \equiv\Bigl\langle {t_+,{\vec x}}
\Bigl\vert {i \over {\cal D} - m_0^2 + i \epsilon} \Bigr\vert {t_-,{\vec 0}}
\Bigr\rangle \; . \eqno(4.10b)$$
With these objects the phase can be reduced to a functional integral over only
metrics and vector potentials:
$$e^{-i \xi} = \int d^3x \; e^{-i {\vec k} \cdot {\vec x}} \Fint [dg] [dA] \;
D[g,A]\Bigl(t_+,{\vec x};t_-,{\vec 0}\Bigr) \; e^{i S_{\rm GR}[g] + i 
S_{\rm EM}[g,A] + i \Gamma_{\phi}[g,A]} \; , \eqno(4.11)$$
\indent Contact is made with particle dynamics by writing the general 
propagator in Schwinger form:
$$\Bigl\langle {t_+,{\vec x}} \Bigl\vert {i \over {\cal D} - m_0^2 +i \epsilon} 
\Bigr\vert {t_-,{\vec 0}} \Bigr\rangle = \int_0^{\infty} ds \; \Bigl\langle 
{t_+,{\vec x}} \Bigl\vert \exp\Bigl[i s ({\cal D} - m_0^2 +i \epsilon)\Bigr]
\Bigr\vert {t_-,{\vec 0}} \Bigr\rangle \; . \eqno(4.12)$$
One then regards the exponent as the Hamiltonian of a first quantized particle
and the expectation value is converted into a functional integral in the usual 
way. We can give this a reparameterization invariant form by regarding the 
proper time as the unfixed part of the einbein $e(\tau)$ in ${\dot e} = 0$ 
gauge:
$$\int_0^{\infty} ds = \Fint [de] \; \delta[{\dot e}] \eqno(4.13)$$
Integrating out the canonical momenta and absorbing any ordering terms into the
measure gives:
$$\eqalignno{\int d^3x\; e^{-i{\vec k} \cdot {\vec x}} \Bigl\langle &{t_+,{\vec 
x}} \Bigl\vert {i \over {\cal D} -m_0^2 +i \epsilon} \Bigr\vert {t_-,{\vec 0}} 
\Bigr\rangle \cr
&= \Fint [de] [d^4\chi] \; \delta[{\dot e}[ \; \delta\Bigl(\chi^0(0)- t_-\Bigr)
\; \delta\Bigl(\chi^0(1) - t_+\Bigr) \; \delta^3\Bigl({\vec \chi}(0)\Bigr) 
&(4.14) \cr
&\qquad \times \exp\Biggl\{-i {\vec k} \cdot {\vec \chi}(1) + i \int_0^1 d\tau 
\Bigl[\frac1{4e} g_{\mu\nu} {\dot \chi}^{\mu} {\dot \chi}^{\nu} -e m_0^2 - q 
{\dot \chi}^{\mu} A_{\mu}\Bigr]\Biggr\}\cr}$$
One now makes the change of variables defined by the reparameterization which
changes the gauge condition to:
$$\chi^0(\tau) \equiv t_- + (t_+ - t_-) \tau \; . \eqno(4.15)$$
The integral over the einbein is done using the functional equivalent of the 
identity:
$$\int_0^{\infty} {dx \over \sqrt{x}} \exp\Bigl[-a x - \frac{b}{x}\Bigr] =
\sqrt{\frac{\pi}{a}} \exp\Bigl[-\sqrt{4 a b}\Bigr] \; . \eqno(4.16)$$
The final form is:
$$\eqalignno{\int &d^3x\; e^{-i{\vec k} \cdot {\vec x}} \Bigl\langle {t_+,{\vec 
x}} \Bigl\vert {i \over {\cal D} - m_0^2 +i \epsilon} \Bigr\vert {t_-,{\vec 0}} 
\Bigr\rangle &(4.17) \cr
&= \Fint [d^3\chi] \delta^3\Bigl({\vec \chi}(0)\Bigr) \exp\Biggl\{-i {\vec k} 
\cdot {\vec \chi}(1) - i \int_0^1 d\tau \Bigl[ m_0 \sqrt{-g_{\mu\nu} {\dot 
\chi}^{\mu} {\dot \chi}^{\nu}} +q {\dot \chi}^{\mu} A_{\mu}\Bigr]\Biggr\}\cr}$$
where $\chi^0(\tau)$ is understood to be defined by (4.15).\hfil\break
\indent Substituting (4.17) into (4.11) gives the following expression for the
phase:
$$\eqalignno{e^{-i \xi} = \Fint [dg] &[dA] [d^3\chi] \; \delta^3\Bigl({\vec 
\chi}(0) \Bigr) \exp\Bigl\{-i {\vec k} \cdot {\vec \chi}(1)\Bigr\} \cr
& \times \exp\Bigl\{i S_{\rm GR}[g] + i S_{\rm EM}[g,A] + i S_{\rm 
part}[g,A,\chi] + i \Gamma_{\phi}[g,A]\Bigr\} \; , \qquad &(4.18a) \cr}$$
where the particle action is:
$$S_{\rm part}[g,A,\chi] \equiv - \int_0^1 d\tau \Biggl[m_0 \sqrt{-
g_{\mu\nu}\Bigl(\chi(\tau)\Bigr) {\dot \chi}^{\mu}(\tau) {\dot \chi}^{\nu}(
\tau)} + q {\dot \chi}^{\mu}(\tau) A_{\mu}\Bigl(\chi(\tau)\Bigr) \Biggr] \; .
\eqno(4.18b)$$
It should be noted again that various ordering corrections have be subsumed 
into the measure. Also note, again, that $\chi^0(\tau)$ is the non-dynamical 
function (4.15).\hfil\break
\line{\hfil}
\line{\hfil}
\line{\bf 5. THE CLASSICAL LIMIT \hfil}
\line{\hfil}
\indent The purpose of this section is to show how the classical limit of $\xi$
relates to the ADM$^{10}$ mechanism discussed in Section 3. This is crucial to 
seeing that the reorganization of of perturbation theory I shall propose in the
next section in fact manifests the gravitational regulation of ultraviolet 
divergences at lowest order.\hfil\break
\indent We can forget about the scalar-induced effective action 
$\Gamma_{\phi}$ because it is a quantum effect. What is necessary for our 
purposes is to solve the classical field equations derived from the 
action:\footnote{*}{The same technique has been used, in the context of 2-body
scattering, by Fabbrichesi, Pettorino, Veneziano and Vilkovisky$^{11}$.}
$$S_{\rm class}[g,A,\chi] = S_{\rm GR}[g] + S_{\rm EM}[g,A] + S_{\rm part}[g,A,
\chi] \; . \eqno(5.1)$$
The boundary conditions for the metric and the vector potential come from the 
asymptotic in and out vacua. Those for the particle are:
$$\chi^i(0) = 0 \qquad , \qquad {\dot \chi}^i(1) = k^i \; . \eqno(5.2a)$$
We can save ourselves a small amount of effort by instead imposing: 
$$\chi^i(0) = 0 \qquad , \qquad {\dot \chi}^i(1) = 0 \; , \eqno(5.2b)$$
and then boosting up to (5.2a). If (5.2b) is used one finds the classical limit
of the phase by evaluating the action at the solution:
$$\xi_{\rm class}(t_+,t_-,0) = -S_{\rm class}[{\widehat g},{\widehat A},
{\widehat \chi}] \; . \eqno(5.3a)$$
One then divides out the time interval and takes the asymptotic limit:
$$m_{\rm class} = \lim_{\scriptscriptstyle t_{\pm} \rightarrow \pm \infty} 
\Bigl({\xi(t_+,t_-,0) \over t_+ - t_-}\Bigr) \; . \eqno(5.3b)$$
\indent Although gravity does regulate this problem, just as it did for that of
ADM,$^{10}$ some of the intermediate expressions will be singular unless the
point particle is smeared out. ADM resolved this issue by converting the 
particle into a spherical shell of radius $\epsilon$ in isotropic coordinates. 
I shall do the same, but I face the additional problem, which they did not, of
keeping the system static for all time. I shall accordingly employ a perfect
fluid regularization in which the point particle is converted into a swarm of 
particles labelled by an internal vector ${\vec \sigma}$:
$$\chi^i(\tau) \longrightarrow X^i(\tau,{\vec \sigma}) \eqno(5.4)$$
The particle's action goes to that of a perfect fluid:
$$S_{\rm part}[g,A,\chi] \longrightarrow - \int d\tau d^3\sigma \Biggl\{\sqrt{-
g_{\alpha \beta} {\dot X}^{\alpha} {\dot X}^{\beta}} \Bigl[\mu({\vec \sigma}) +
{\Pi({\vec \sigma}) \over \sqrt{-g}} \Bigr] + {q \over m_0} \mu({\vec \sigma})
{\dot X}^{\mu} A_{\mu}\Biggr\} \; , \eqno(5.5)$$
with number density $n(x)$ given by $\mu({\vec \sigma})$:
$$n(x) = {1 \over \sqrt{-g}} \int d\tau d^3\sigma \; {\mu({\vec \sigma}) \over 
m_0} \delta^4\Bigl(x - X(\tau,{\vec \sigma})\Bigr) \sqrt{-g_{\alpha \beta} 
{\dot X}^{\alpha} {\dot X}^{\beta}} \; ,\eqno(5.6a)$$
and pressure $p(x)$ given by $\Pi({\vec \sigma})$:
$$p(x) =- {1 \over g} \int d\tau d^3\sigma \; \Pi({\vec \sigma}) \; 
\delta^4\Bigl(x - X(\tau,{\vec \sigma})\Bigr) \sqrt{-g_{\alpha \beta} 
{\dot X}^{\alpha} {\dot X}^{\beta}} \; ,\eqno(5.6b)$$
I shall follow ADM in taking the mass density to be that of a spherical shell:
$$\mu({\vec \sigma}) \equiv {m_0 \delta(\sigma - \epsilon) \over 4 \pi 
\epsilon^2} \; , \eqno(5.7a)$$
however, I shall impose a negative internal pressure:
$$\Pi({\vec \sigma}) \equiv - f(\epsilon) \theta(\epsilon - \sigma) \; ,
\eqno(5.7b)$$
to hold the shell together. Duff has shown that this does not affect the ADM
mass.$^{12}$ The function $f(\epsilon)$ is a non-dynamical constant to be 
determined shortly.\hfil\break
\indent The manifest spherical symmetry and the assumed time translation 
invariance of this problem suggest that we look for a solution of the form:
$${\widehat g}_{\mu\nu} dx^{\mu} dx^{\nu} = -A^2(r) dt^2 + B^2(r) d{\vec x} 
\cdot d{\vec x} \; , \eqno(5.8a)$$
$${\widehat A}_{\mu} dx^{\mu} = A_0(r) dt \; , \eqno(5.8b)$$
$${\widehat X}^{\mu}(\tau,{\vec \sigma}) = \delta^{\mu}_0 \Bigl(t_- + (t_+ - 
t_-) \tau \Bigr) + \delta^{\mu}_i \sigma^i \; . \eqno(5.8c)$$
The solution for $A(r)$ has the form $A(r) = \alpha(r)/B(r)$ with:
$$\alpha(r) = \theta(r-\epsilon) \Bigl[1 + {Q^2 - M^2 \over 4 r^2}\Bigr] +
\theta(\epsilon - r) \Bigl[1 + \Bigl({Q^2 - M^2 \over 4 \epsilon^2}\Bigr) 
\Bigl(2 - {r^2 \over \epsilon^2}\Bigr)\Bigr] \; . \eqno(5.9a)$$
The two other functions work out to be:
$$B(r) = \theta(r - \epsilon) \Bigl[1 + {M \over r} - \Bigl({Q^2 - M^2 \over 4 
r^2}\Bigr)\Bigr] + \theta(\epsilon - r) \Bigl[1 + {M \over \epsilon} - 
\Bigl({Q^2 - M^2 \over 4 \epsilon^2}\Bigr)\Bigr] \; , \quad \eqno(5.9b)$$
$$A_0(r) = {q \over 4 \pi B(r)} \Bigl\{ {\theta(r - \epsilon) \over r} +
{\theta(\epsilon - r) \over \epsilon}\Bigr\} \; . \eqno(5.9c)$$
The parameters in relations (5.9a-c) have been represented as lengths according
to the standard convention of geometrodynamics:
$$M_0 \equiv G m_0 \qquad , \qquad Q \equiv q \sqrt{G \over 4 \pi} \; ,
\eqno(5.10a)$$
$$M \equiv \epsilon \Biggl[-1 + \sqrt{1 + {2 M_0 \over \epsilon} + {Q^2 \over
\epsilon^2}} \Biggr] \; . \eqno(5.10b)$$
The necessary internal pressure constant $f(\epsilon)$ works out to be:
$$f(\epsilon) = {B^3(\epsilon) \over 8 \pi G} \Bigl({Q^2 - M^2 \over 
\epsilon^4}\Bigr) \; . \eqno(5.11)$$
\indent Since the solution is static the action must be a constant multiplied 
by the time interval. However, this constant turns outs not to be minus the ADM
mass but rather a sort of enthalpy reflecting the presence of the pressure in 
the perfect fluid regularization of the particle action:
$$\eqalignno{\Bigl(S_{\rm GR} + S_{\rm EM} + S_{\rm part}\Bigr)\Bigl\vert_{\rm 
solution} &= - {M \over G} (t_+ - t_-) - \int d^4x \sqrt{-g} \; p \; ,
&(5.12a) \cr
&\equiv -\Bigl(U + p V\Bigr) (t_+ - t_-) \; . &(5.12b) \cr}$$
The energy $U$ and the $p V$ term can be evaluated for any $\epsilon$. They 
have the following simple forms:
$$U = {M \over G} \qquad , \qquad p V = -{1 \over 3 G} (M - M_0) \; .
\eqno(5.13)$$
In the limit $\epsilon \rightarrow 0$ the energy just gives the ADM mass (3.4).
The $p V$ term remains finite in this limit, but neither does it vanish. Its 
physical interpretation seems to be that gravity is not sufficient to hold the 
charge together. This means the calculation is not really consistent. I shall 
do better shortly, but do not let this obscure two important facts:
\item{(1)} Contact has been established, modulo the $p V$ term, between the 
standard quantum field theoretic definition of mass and the classical ADM 
calculation of the self-energy of a gravitating, point charged particle.
\item{(2)} Even with the $p V$ term, gravity has suppressed what would 
otherwise be a divergent result.

\line{\hfil}
\line{\hfil}
\line{\bf 6. QUANTUM MECHANICAL INTERPRETATION \hfil}
\line{\hfil}
\indent The purpose of this section is to introduce the promised reorganization
of conventional perturbation theory. The starting point is the expression 
(4.18a) obtained for the phase at the end of Section 4:
$$\eqalignno{e^{-i \xi} = \Fint [dg] &[dA] \; \exp\Bigl\{i S_{\rm GR}[g] + 
i S_{\rm EM}[g,A] + i \Gamma_{\phi}[g,A]\Bigr\} \cr
&\times \Fint [d\chi] \; \delta^3\Bigl({\vec \chi}(0)\Bigr) \exp\Bigl\{-i {\vec
k} \cdot {\vec \chi}(1) + i S_{\rm part}[g,A,\chi]\Bigr\} \; . &(6.1) \cr}$$
The second line of this expression can be interpreted as the amplitude for a
quantum mechanical particle to go from a delta function at $t = t_-$:
$$\psi_-({\vec x}) = \delta^3({\vec x}) \; , \eqno(6.2a)$$
to a plane wave at $t = t_+$:
$$\psi_+({\vec x}) = e^{i {\vec k} \cdot {\vec x}} \; , \eqno(6.2b)$$
Let us denote the associated action as:
$$\exp\Bigl\{i S_{\rm prop}[g,A]\Bigr\} \equiv \Fint [d\chi] \; \delta^3\Bigl(
{\vec \chi}(0)\Bigr) \exp\Bigl\{-i {\vec k} \cdot {\vec \chi}(1) + i
S_{\rm part}[g,A,\chi]\Bigr\} \; . \eqno(6.3)$$
I define the 0-th order term in the reorganized perturbation theory to be the
stationary phase approximation to the functional integral over metrics and
vector potentials with the following action:
$$S_{\rm class}[g,A] = S_{\rm GR}[g] + S_{\rm EM}[g,A] + S_{\rm prop}[g,A] \; .
\eqno(6.4)$$
\indent To see what diagrams this term captures it is simplest to identify the
ones it misses. No closed scalar loops are included since the scalar-induced 
effective action $\Gamma_{\phi}$ was excluded from (6.4). This does not mean 
there are no scalar lines at all. Owing to the presence of $S_{\rm prop}$ there
must be a single, continuous scalar line in all the included diagrams. Any 
number of graviton and gauge lines can be attached to this line. However, the 
restriction to stationary phase means that we include no closed gauge and/or 
graviton loops which do not include some portion of the single, continuous 
scalar line. {\it So the 0-th order term I have defined consists of all 
diagrams with a single, continuous scalar line and no closed loops which do not
include this line.} One can imagine that the next order term consists of 
diagrams with one closed loop external to the in-out line, and so on.
\hfil\break
\indent It is obvious from the way I have defined it that the 0-th order term
constitutes a gauge invariant resummation of an infinite subset of diagrams. It
is also obvious this 0-th term contains the classical limit considered in the 
previous section. Further, it will be rendered {\it less} singular, not more,
by the inevitable quantum spread of the particle. All of this implies that the
0-th term just defined must manifest the gravitational suppression of 
divergences.\hfil\break
\indent The physical interpretation of the 0-th order approximation to $\xi$ is
the phase developed by a quantum mechanical particle moving in the potentials
generated by its own probability current. Whether or not there is any chance of
being able to compute it depends upon which of the following two possibilities
is realized:
\item{(1)} The particle cannot form bound states in its own potentials; or
\item{(2)} The particle can form bound states in its own potentials.

\noindent In case (1) we are left with a complicated, time dependent scattering
problem which seems to be intractable. However, many simplifications are
possible in case (2).\hfil\break
\indent If bound states form one can forget about the asymptotic wavefunctions 
(6.2a-b). In the limit of infinite time separation the phase will be dominated 
by the lowest energy state. Further, one need only compute $S_{\rm prop}[g,A]$ 
for a class of metrics and vector potentials which is broad enough to include
the eventual solution. In the scalar problem we could immediately reduce from
nine functions of $x^{\mu}$ to a static, spherically symmetric system 
characterized by only three functions of a single variable:
$$g_{\mu\nu} dx^{\mu} dx^{\nu} = -A^2(r) dt^2 + B^2(r) d{\vec x} \cdot 
d{\vec x} \eqno(6.5a)$$
$$A_{\mu} dx^{\mu} = A_0(r) dt \eqno(6.5b)$$
Finally, variational methods can be usefully applied. If one simply guesses the
wavefunction, assuming static potentials, and then minimizes the total energy, 
the result will be an upper bound on the true 0-th order mass. Note, in this 
context, that {\it any} finite result would be awe inspiring.\hfil\break
\line{\hfil}
\line{\hfil}
\line{\bf 7. QUANTUM MECHANICS IN REISSNER-NORDSTROM \hfil}
\line{\hfil}
\indent The purpose of this section is to ascertain which of the two cases 
pertains to a charged, gravitating scalar: can it bind to its own potentials or
not? We can immediately specialize to the static, spherically symmetric 
potentials (6.5a-b). Modulo the effects of operator ordering, the Hamiltonian 
is:
$$H = {A(r) \over B(r)} \sqrt{\Vert {\vec p} \Vert^2 + m_0^2 B^2(r)} + q A_0(r)
\eqno(7.1)$$
\indent Assuming the particle is bound, we can invoke Birkhoff's theorem to
fix the potentials outside most of the particle's probability density:
$$A_{\rm ext}(r) = {1 \over B_{\rm ext}} \Biggl[1 + \Bigl({Q^2 - M^2 \over 
4 r^2}\Bigr) \Biggr] \; , \eqno(7.2a)$$
$$B_{\rm ext}(r) = 1 + {M \over r} - \Bigl({Q^2 - M^2 \over 4 r^2}\Bigr) \; ,
\eqno(7.2b)$$
$$A_0^{\rm ext}(r) = {q \over 4 \pi} {1 \over r B_{\rm ext}} \; . 
\eqno(7.2c)$$
The various parameters have been expressed as lengths in the usual 
geometrodynamical convention:
$$Q \equiv q \sqrt{G \over 4 \pi} \qquad , \qquad M_0 \equiv m_0 G \; , 
\eqno(7.3)$$
however, it should be noted that $M$ is at this stage undetermined. We can also
assume that the momentum is dominated by uncertainty pressure:
$$p \sim {\hbar \over r} \qquad \Longrightarrow P \equiv G \; p \sim {L_P^2 
\over r} \; , \eqno(7.4)$$
where $L_P$ is the Planck length.\hfil\break
\indent If we geometrodynamicize the Hamitonian (${\cal H} \equiv G H)$, then 
its form beyond most of the probability density is:
$${\cal H}_{\rm ext} = {A_{\rm ext} \over B_{\rm ext}} \sqrt{P^2 + 
M_0^2 B^2_{\rm ext}} + {Q^2 \over r B_{\rm ext}} \eqno(7.5)$$
At large $r$ this becomes:
$${\cal H}_{\rm ext} \sim M_0 + (Q^2 - M_0 M) {1 \over r} \qquad r \gg Q
\eqno(7.6)$$
One consequence of (7.3) is that the ratio $Q/L_P$ goes like the square root of
the fine structure constant, whereas the $M_0/L_P$ is the ratio of $m_0$ to the
Planck mass. It follows that any particle which is relevant to low energy 
physics must obey:
$$M_0, M \ll Q \eqno(7.7)$$
In this case we see that the Hamiltonian falls off asymptotically, suggesting
that no reasonably light bound state can form.\hfil\break
\indent It is not really consistent to use the external potentials in the 
interior but doing so fails to reveal an inner region of binding. In isotropic 
coordinates the singularity occurs at $r_0 = (Q - M)/2$. Specializing to a 
point just slightly outside gives only another repulsive Hamiltonian:
$${\cal H}_{\rm ext} \sim {(Q - M)^2 L_P^2 \over 4 Q \epsilon^2} \qquad r =
\frac12 (Q - M) + \epsilon \eqno(7.8)$$
\indent It seems fair to conclude that any charged scalar bound states would
necessarily have Planck scale masses. On the other hand, setting $Q = 0$ in 
(7.6) seems to suggest that the chargeless scalar can form a bound state. 
Expression (7.8) suggests that quantum uncertainty pressure protects it from
collapse, unlike the neutral scalar studied classically by ADM.$^{10}$
\hfil\break
\line{\hfil}
\line{\hfil}
\line{\bf 8. DISCUSSION \hfil}
\line{\hfil}
\indent I have proposed a gauge invariant reorganization of conventional 
perturbation theory in which gravitational regularization is a 0-th order 
effect. The existence of any new technique deserves comment because it might 
be thought that the possibilities for one have been pretty well exhausted by 
now. A new expansion must be in terms of some parameter, such as the dimension 
of spacetime$^{13}$ or the number of matter fields,$^{14}$ and there simply 
aren't any plausible parameters that have not been tried.\hfil\break
\indent The secret of my expansion is that it does not conform to the usual
rules which require the parameter to appear in the Lagrangian. I have instead 
exploited a parameter which depends, to some extent, on the thing being 
computed. This parameter is the number of closed loops which are external to 
continuous matter lines that come in from the asymptotic past and proceed out
to the asymptotic future. Not all processes have such lines. However, the 
technique can be used on those that do, and any evidence for the 
non-perturbative viability of quantum General Relativity would be interesting.
\hfil\break
\indent Of particular interest are the self-energies of matter particles. The
classical computation of ADM,$^{10}$ summarized in Section 3, suggests that 
this is a natural setting for conventional perturbation theory to break down. 
If quantum gravity regulates ultraviolet divergences as classical gravity
certainly does then the asymptotic expansion must contain inverse powers or
logarithms.\hfil\break
\indent I was able to reexpress the standard definition of the pole of the 
propagator in terms of the new expansion. The 0-th order term has the physical 
interpretation of the phase developed by a quantum mechanical particle moving 
in the potentials induced by its own probability current. If these potentials 
cannot form bound states one has an intractable scattering problem. However, 
the 0-th term is eminently calculable if there are bound states. In this case 
the lowest energy state dominates. One can also assume that the potentials are 
static, and that they possess simplifying symmetries. If nothing else works, it 
is always possible to obtain an upper bound on the mass through variational
techniques.\hfil\break
\indent The explicit analysis of Section 7 indicates that there is probably not
a charged bound state scalar of less than about the Planck mass. However, it
does seem possible that light neutral scalars can form. Adding spin complicates
the gravitational and electrodynamic potentials enormously. It also adds a new
parameter in the form of the spin-to-mass ratio $a$. It may be very significant
that, whereas the charge parameter completely dominates the mass, the 
spin-to-mass ratio is larger by almost the same ratio. For an electron one 
finds:
$$M \sim 10^{-55}~{\rm cm} \ll -Q \sim 10^{-34}~{\rm cm} \ll a \sim 10^{-11}~{
\rm cm} \; , \eqno(8.1)$$
so it is not unreasonable to expect large spin-dependent forces. The physical 
interpretation for this is that different portions of a rapidly spinning body 
see one another through enormous relative boosts. What is a minuscule matter
density in our frame can therefore seem overwhelming from the instantaneous 
rest frame of a spinning observer.\footnote{*}{I wish to thank D. N. Page for
elucidating this point.} So there is some hope for getting light fermionic
bound states.\hfil\break
\line{\hfil}
\line{\hfil}
\line{\bf ACKNOWLEDGEMENTS \hfil}
\line{\hfil}
\indent It is a pleasure to acknowledge almost twenty years of conversations on
this subject with S. Deser and N. C. Tsamis. This work was partially supported
by DOE contract DE-FG02-97ER41029 and by the Institute for Fundamental Theory.
\line{\hfil}
\line{\hfil}
\line{\bf REFERENCES \hfil}
{\tenpoint
\line{\hfil}
\noindent ~1. G. `t Hooft and M. Veltman, {\sl Ann. Inst. Henri Poincar\'e} 
{\bf 20} (1974) 69.\hfil\break
\noindent ~2. M. Goroff and A. Sagnotti, {\sl Phys. Lett.} {\bf 106B} (1985) 
81; \hfil\break
\indent {\sl Nucl. Phys.} {\bf B266} (1986) 709.\hfil\break
\noindent ~3. S. Deser and P. van Nieuwenhuizen, {\sl Phys. Rev.} {\bf D10}
(1974) 401.\hfil\break
\noindent ~4. S. Deser and P. van Nieuwenhuizen, {\sl Phys. Rev.} {\bf D10}
(1974) 411.\hfil\break
\noindent ~5. S. Deser, Hung Sheng Tsao and P. van Nieuwenhuizen, {\sl Phys.
Rev.} {\bf D10} (1974) 3337.\hfil\break
\noindent ~6. S. Deser, {\sl Rev. Mod. Phys.} {\bf 29} (1957) 417.\hfil\break
\noindent ~7. B. S. DeWitt, {\sl Phys. Rev. Lett.} {\bf 13} (1964) 114.
\hfil\break
\noindent ~8. I. B. Khriplovich, {\sl Soviet J. Nucl. Phys.} {\bf 3} (1966) 
415.\hfil\break
\noindent ~9. C. J. Isham, A. Salam, and J. Strathdee, {\sl Phys. Rev.} {\bf 
D3} (1971) 1805.\hfil\break
\noindent 10. R. Arnowit, S. Deser and C. W. Misner, {\sl Phys. Rev. Lett.} 
{\bf 4} (1960) 375;\hfil\break
\indent {\sl Phys. Rev.} {\bf 120} (1960) 313;\hfil\break
\indent {\sl Phys. Rev.} {\bf 120} (1960) 321;\hfil\break
\indent {\sl Ann. Phys.} {\bf 33} (1965) 88.\hfil\break
\noindent 11. M. Fabbrichesi, R. Pettorino, G. Veneziano and G. A. Vilkovisky, 
{\sl Nucl. Phys.} {\bf B419} (1994) 147.\hfil\break
\noindent 12. M. J. Duff, {\sl Phys. Rev.} {\bf D7} (1973) 2317.\hfil\break
\noindent 13. A. Strominger, {\sl Phys. Rev.} {\bf D24} (1982) 3082.\hfil\break
\noindent 14. A. Strominger, A gauge invariant resummation of quantum gravity, 
in: {\it International Symposium}\hfil\break
\indent {\it on Gauge Theory and Gravitation, Nara, Japan, Aug. 20-24, 1982}, 
eds., K. Kikkawa, N. Nakan-\hfil\break
\indent ishi and H. Nariai, Springer-Verlag, (1983).\hfil}
\bye